\documentclass[eqsecnum,twocolumn,citeautoscript]{revtex4}
\usepackage{bm}
\usepackage{graphicx}
\usepackage{amsmath}

\newcommand{\figuresatend}{}

\newcommand{\TIJ}{T(I \to J)}
\newcommand{\TJI}{T(J \to I)}
\newcommand{\TIJi}{T_\infty(I \to J)}
\newcommand{\TJIi}{T_\infty(J \to I)}
\newcommand{\TIJe}{\widetilde{T}_\infty(I \to J)}
\newcommand{\TJIe}{\widetilde{T}_\infty(J \to I)}
\newcommand{\tij}{t(i \to j)}
\newcommand{\tji}{t(j \to i)}
\newcommand{\paccij}{P_\text{acc}(i \to j)}

\newcommand{\aij}{\alpha(i \to j)}
\newcommand{\aji}{\alpha(j \to i)}
\newcommand{\diI}{\delta_{iI}}
\newcommand{\djJ}{\delta_{jJ}}

\newcommand{\dosabs}{\Omega}
\newcommand{\dos}[1]{\dosabs(#1)}

\begin{document}

\makeatletter
\floats@sw{\newcommand{\mycaption}[1]{\caption{#1}}}
{\setlength{\abovecaptionskip}{2 in}\newcommand{\mycaption}[1]{\caption[#1]{Shell, Debenedetti, and Panagiotopoulos}}}
\renewcommand{\l@figure}[2]{\figurename\ #1\\[2cm]}
\makeatother

\title{An improved Monte Carlo method for direct calculation of the density of states}

\author{M. Scott Shell}
\email[email:]{shell@princeton.edu}
\author{Pablo G. Debenedetti}
\email[email:]{pdebene@princeton.edu}
\author{Athanassios Z. Panagiotopoulos}
\email[email:]{azp@princeton.edu}
\affiliation{Department of Chemical Engineering\\Princeton University, Princeton, NJ 08544}
\date{July 16, 2003}

\begin{abstract}
We present an efficient Monte Carlo algorithm for determining the density of states which is based on the statistics of transition probabilities between states.  By measuring the infinite temperature transition probabilities\textemdash that is, the probabilities associated with move \emph{proposal} only\textemdash we are able to extract excellent estimates of the density of states.  When this estimator is used in conjunction with a Wang-Landau sampling scheme [F. Wang and D.~P. Landau, Phys. Rev. Lett. \textbf{86}, 2050 (2001)], we quickly achieve uniform sampling of macrostates (e.g., energies) and systematically refine the calculated density of states.  This approach requires only potential energy evaluations, continues to improve the statistical quality of its results as the simulation time is extended, and is applicable to both lattice and continuum systems.  We test the algorithm on the Lennard-Jones liquid and demonstrate good statistical convergence properties.
\end{abstract}

\maketitle

\section{Introduction}

In the fifty years since it was first introduced \cite{metropolis1}, the Monte Carlo (MC) method has become one of the primary tools for the simulation of materials and the prediction of their properties.  MC simulations are widely used and a significant number of methodological improvements now exist for efficient and accurate studies of systems which are otherwise intractable with the traditional Metropolis sampling \cite{allentildesley1,frenkelsmit1}.  Examples include high-density or low-temperature liquids, network-forming fluids, polymers, and proteins, for which the simulated system is prone to becoming trapped in potential energy minima for large numbers of simulation steps.  

A number of these extended MC methods solve the ergodicity problem by forcing a broad sampling of phase space; included in this category are parallel tempering \cite{paralleltemp}, multicanonical methods \cite{berg1}, and the more recent Wang-Landau (WL) algorithm \cite{wanglandau1}.  The latter two achieve broad phase space sampling by attempting to produce a uniform distribution of one or more macroscopic observables, such as number of particles, volume, and potential energy.  To arrive at such a distribution, each microstate of the system must be sampled with a probability inversely proportional to the degeneracy of the corresponding macroscopic observable.  Thus whether implicitly (as part of a state weight) or explicitly (through direct calculation), these methods must determine the density of states, $\Omega$.  Of course, $\Omega$ contains a complete thermodynamic description of the system via Boltzmann's equation,
\begin{equation}
S(N,V,E) = k \ln{\Omega(N,V,E)}.
\label{boltzmanns}
\end{equation}
Clearly, the density of states is not known prior to the simulation but must be successively approximated during its course.  The multicanonical approach iteratively updates its estimate for weights (which include $\Omega$) from histogram results of several pre-production runs; once the density of states is reasonably converged, it and histogram data from a final long simulation interval are ``reweighted'' to determine properties at the state conditions of interest \cite{berg1,wilding1}.   In the WL scheme, microstates are explicitly sampled with the probability $\Omega^{-1}$ and the density of states is modified at every simulation step \cite{wanglandau1,shell1}; for each visited state, the corresponding value of $\Omega$ is scaled by the modification factor, $f$.  In this way, the the density of states is designed to converge to its correct value when macroscopic properties are sampled uniformly.  However, to eliminate $f$ fluctuations in the final result, a schedule must be imposed in which $f$ changes from an initially large value (say, $f \approx e$) to nearly unity.  Often, this schedule is implemented by maintaining a histogram of visited states, and when the bin with the fewest entries is no less than $80\%$ of the average bin (i.e., when the histogram is sufficiently flat), $f$ is changed according to $f_\text{new}=\sqrt{f_\text{old}}$ and the histogram is reset.

The WL algorithm has received much attention lately, perhaps due to its straightforward implementation and wide applicability, and a number of methodological variants have been proposed, including multibondic \cite{wlmultibondic}, $N$-fold way \cite{wlnfoldway}, and quantum \cite{wlquantum} extensions.  Of relevance to continuum systems, Yan and de Pablo \cite{jjdp2} noticed two deficiencies of the WL method: estimates of the density of states reach a limiting statistical accuracy which is not improved with further MC steps, and the large number of configurations generated towards the end of the simulation make only a small contribution to the calculated density of states.  They proposed two variants in which estimates of temperature improve the convergence of the density of states relative to the original method \cite{jjdp2}.  Both of these, however, must be formulated specifically for the relevant macroscopic parameter (the authors originally discussed only energy in this regard).  Furthermore, their configurational temperature method requires first and second derivatives of the potential energy, which limits it to systems with continuous potentials and, practically speaking, excludes those for which such calculations are computationally expensive.   

Here we present a MC implementation for directly determining the density of states of a system, which we believe is superior to original WL algorithm while retaining much of its generality.  A reformulation of the Wang-Landau approach, this implementation is based on the ideas of transition matrix MC methods: the calculations utilize a record of proposed transitions between, rather than visits to, various macrostates \cite{smithbruce1, smithbruce2, oliveira1, fitz1, fitz2, wang1, wang2, wang3, wang4, wang5, jre1, jre2}.  Our method builds on the ideas of Ref.\ \onlinecite{jjdp2} in that we supplement the normal WL sampling scheme with this additional data collection, which ultimately produces density of states estimates of higher statistical quality than the WL estimates.  In this paper we show that our method is: (1) general\textemdash applicable to lattice and continuum systems; (2) robust\textemdash continuously improves its estimate for the density of states, regardless of a particular modification factor schedule; (3) simple\textemdash requires only potential energy evaluations; (4) efficient\textemdash uses all collected information during the course of a simulation; and (5) accurate\textemdash produces density of states estimates of higher statistical quality than the WL method for a given number of simulation steps.  

In what follows we first outline the theoretical basis for transition matrix calculations and review a previous method based wholly on transition probabilities \cite{oliveira1,wang3,wang4,wang5}.  We then present our implementation, which combines transition matrix ideas with Wang-Landau sampling, and discuss its advantages relative to both the WL and transition matrix formulations alone.  Finally, we proceed to discuss results for a set of test cases.

\section{Theoretical basis}

We begin by considering the transition probability of moving from macrostate $I$ to $J$ given that the system is currently in $I$, which we denote by $\TIJ$.  This derivation applies generally to any well-defined macrostate; in this work, however, we use $I$ and $J$ to denote energy levels.  (We refer the reader to Refs.\ \onlinecite{jre1} and \onlinecite{jre2} which demonstrate the use of transition probabilities with a different macrostate variable.)  The expression for $\TIJ$ in terms of the corresponding microstate transition probabilities is then
\begin{equation}
\TIJ = \dos{I}^{-1} \sum_i \sum_j \tij \, \diI \, \djJ 
\label{basiceqn}
\end{equation}
where the sums extend over all microstates, $\dos{I}$ gives the degeneracy of macrostate $I$, $\tij$ is the probability that the system will make a move from microstate $i$ to $j$ given that it is currently in state $i$, and $\diI$ (equivalently, $\djJ$) is defined such that
\begin{equation}
\diI = \left\{
\begin{array}{ll} 1 & \text{if } i \in I \\ 0 & \text{otherwise}. \end{array}
\right.
\end{equation}
That is, $\diI$ is one if and only if microstate $i$ is a member of macrostate $I$.  Implicit in Eq.\ \ref{basiceqn} is the assumption of equal \emph{a priori} probability of microstates; the macrostate transition probability is simply the microcanonical average of all corresponding microstate transition probabilities.  Accordingly, the following results apply to any simulation procedure which generates microcanonical averages according to Eq.\ \ref{basiceqn}.  By dividing Eq.\ \ref{basiceqn} by the corresponding expression for the reverse transition, we obtain an important relation between the transition probabilities and density of states:
\begin{equation}
\frac{\TIJ}{\TJI} = \frac{\dos{J}}{\dos{I}} \left( \frac{\sum_i \sum_j \tij \, \diI \, \djJ}{\sum_i \sum_j \tji \, \diI \, \djJ} \right).
\label{ratio}
\end{equation}
The microstate transition probability $\tij$ appearing in these expressions is the product of two terms: $\aij$, the probability of proposing a transition from $i$ to $j$, dependent only on the type of MC move employed; and $\paccij$, the probability of accepting a proposed move from $i$ to $j$, dependent on the state conditions of the particular statistical ensemble.  We consider the special case in which moves are strictly accepted, $\paccij=1$, which corresponds to canonical Metropolis-style sampling at infinite temperature \cite{inftcomment}.  Consequently, we denote $\TIJi$ as that corresponding to strict move acceptance and will refer to it as an infinite temperature transition probability.  We note that $\paccij=1$ is not necessarily the infinite-$T$ limit for all ensembles and classes of simulation moves; however, in identifying $\TIJi$ with the strict acceptance condition, this derivation is in fact quite general, regardless of the particular move type or ensemble used.  The relationship among the infinite-temperature macrostate transition probabilities is then
\begin{equation}
\frac{\TIJi}{\TJIi} = \frac{\dos{J}}{\dos{I}} \left( \frac{\sum_i \sum_j \aij \, \diI \, \djJ}{\sum_i \sum_j \aji \, \diI \, \djJ} \right).
\label{transdoseqn}
\end{equation}
Given the symmetry properties of a particular move type [i.e., the ratio $\aij/\aji$], this equation permits an estimate of the density of states from knowledge of macrostate transition probabilities.  In the case of symmetric moves with $\aij=\aji$, the term in parenthesis in Eq.\ \ref{transdoseqn} drops out completely; examples of such moves include single-spin flips and single-particle displacements.  It is important to consider $\TIJi$ as the probability that a move will be \emph{proposed} from $I$ to $J$.  Thus, the collection of $\TIJi$ quantities gives a measure of the connectivity of macrostates for a given move type.  Previous studies have demonstrated that this type of information is statistically superior to histogram data \cite{smithbruce1,fitz1,wang5}, and its use is central to our method.

Though the preceding derivation entails discrete macrostates, the extension to off-lattice systems is conceptually identical and thus relatively straightforward \cite{wang5}.  Practically speaking, however, one must impose a discretization scheme on any continuous macroscopic variables for the measurement of transition probabilities in a simulation.  Because of this discretization, the microstate sampling scheme must also be discretized.  For example, each configuration's energy must be assigned an effective value, the average energy of the corresponding bin, for use in the acceptance criterion.  Lack of congruence with the transition matrix discretization violates the microcanonical average in Eq.\ \ref{basiceqn} and results in systematic errors in the density of states. 

\section{Algorithmic details}

To calculate $\dosabs$ for a given system, two algorithmic ingredients are necessary, as discussed by Oliveira, et. al. \cite{oliveira1}: (i) a good way to measure macrostate degeneracies, termed an ``estimator,'' and (ii) broad sampling of phase space such that all macrostates of interest are visited.  The first condition is met by measurement of macrostate transition probabilities and use of Eq.\ \ref{transdoseqn}.  One can estimate infinite-$T$ transition probabilities by recording move \emph{proposal} statistics during the course of a simulation, even though the acceptance of the moves may not correspond with the infinite-$T$ condition \cite{wang3,wang4,wang5}.  In other words, the $\TIJi$ values can be determined using any arbitrary sampling scheme for $\paccij$ since they depend only on the proposal probabilities contained in $\aij$.  The procedure involves adding $1$ to a matrix entry $C(I,J)$ every time a move from macrostate $I$ to $J$ is proposed.  The estimate for the transition probability is
\begin{equation}
\TIJe = C(I,J) / \sum_K C(I,K)
\label{transapprox}
\end{equation}
where sum extends over all macrostates and the tilde indicates the estimate.  Once the infinite-$T$ transition probabilities are known, the density of states can be calculated from Eq.\ \ref{transdoseqn}.  In general, this is an over-specified problem since in a system of $N$ macrostates with $N$ unknown values of $\dosabs$ (more accurately, $N-1$ unknown \emph{relative} values), there are $N(N-1)/2$ such equations.  In solving for the density of states given the transition probabilities, the easiest approach considers only neighboring states for $N-1$ instances of Eq.\ \ref{transdoseqn}.  However, one can consider all $N(N-1)/2$ equations by minimizing the variance among the predicted values of $\ln{\Omega}$, as described by Ref.\ \onlinecite{wang5}.  Denoting $S(I)=\ln{\dos{I}}$ and $H(I)=\sum_K C(I,K)$, one minimizes the total variance with respect to the values $S(I)$, where the variance is defined as
\begin{equation}
\sigma_\text{tot}^2 = \sum_{I,J} \frac{[S(I) - S(J) + \ln{\TIJi / \TJIi} ]^2}{\sigma_{IJ}^2}
\label{eq:varminimize}
\end{equation}
with 
\begin{equation}
\sigma_{IJ}^2 = C(I,J)^{-1} + H(I)^{-1} + C(J,I)^{-1} + H(J)^{-1}.
\label{eq:varentry}
\end{equation}
Here, we use the number of entries in the $C$ matrix to determine the weight of each estimate in the minimization procedure, assuming that $\text{var}[C(I,J)] \propto C(I,J)$.  In the solution to these equations, one value of the density of states must be fixed and the remaining can be solved by matrix inversion.

The second condition for calculation of the density of states, broad phase space sampling, can be accomplished by using a uniform ensemble in which all macrostates are equiprobable.  In such an ensemble, the probability of a given microstate is inversely proportional to the density of states of the corresponding macrostate.  Accordingly, the acceptance criterion for symmetric moves is
\begin{equation}
\paccij = \text{min}\left(1,\frac{\dos{I}}{\dos{J}}\right) .
\label{acccriterion}
\end{equation}
Of course, the density of states which comes into play in this expression is not known \emph{a priori}.  A straightforward implementation of transition probabilities is possible by combining Eq.\ \ref{transdoseqn} with this acceptance criterion \cite{wang5}:
\begin{equation}
\paccij = \text{min}\left(1,\frac{\TJIe}{\TIJe}\right) .
\label{acccriterion2}
\end{equation}
With Eqs.\ \ref{transapprox} and \ref{acccriterion2}, one can immediately construct a complete MC procedure, an idea originally proposed by Oliveira, et. al. \cite{oliveira1} and later generalized by Wang and coworkers \cite{wang3,wang4,wang5}.  The simulation is started with $C(I,J)=0$ for all $I$ and $J$.  Moves are then proposed and accepted/rejected based on the acceptance criterion in Eq.\ \ref{acccriterion2}, and after each such step, the $C$ matrix is updated to reflect the proposed move, regardless of whether or not it was accepted.  In this fashion, the approximations $\TIJe$ grow more accurate and the distribution of sampled macrostates becomes increasingly more uniform.  Here, it is important to note that even though the actual acceptance probability is not that in the infinite-$T$ case, the estimates for $\TIJi$ are constructed as if $\paccij=1$ due to the way the $C$ matrix is updated.  The density of states is never directly referenced in this scheme.  Instead, one is concerned only with the infinite-$T$ transition probabilities, and therefore need only maintain the $C$ matrix.  At the end of the simulation or at any intermediate point, an estimate for density of states can be determined using Eq.\ \ref{transdoseqn}.  This estimate improves as the number of steps in the simulation increases, and furthermore, the quality of the estimate can be assessed by the number of entries in each $C(I,J)$ \cite{smithbruce1}.  In this paper, we term this approach the transition matrix (TM) method.

The procedure just described is essentially an iterative scheme for the solution of a set of nonlinear equations describing the density of states \cite{wang3,wang4,wang5}.  In general, this is a problem for which convergence is not immediately obvious nor guaranteed.  In particular, problems arise with the acceptance criterion in Eq.\ \ref{acccriterion2} early on in the simulation when numerous zeros exist in the $C$ matrix; an estimate $\TIJe$ is ill-defined if either $C(I,J)=0$ or $\sum_K C(I,K)=0$.  Consequently, for runs which probe large ranges of $\Omega$, the time required for all macrostates to be initially visited can become extraordinarily long.  In contrast, the Wang-Landau approach forces the complete set of macrostates to be visited very quickly at the beginning of the simulation, but lacks the accuracy and continuous improvement in calculating the density of states that the TM method exhibits.

Fortunately, one can combine the strengths of both algorithms, an implementation that we denote as the WL-TM method.  In this approach, we use the original acceptance criterion in Eq.\ \ref{acccriterion} and update the density of states every move as in the WL method, using $\dos{I} \gets f \times \dos{I}$ for the state at the end of each MC step.  In addition, we start with a zeroed $C$ matrix, as in the TM method, and record all proposed transitions during the entire schedule of modification factors (i.e., the $C$ matrix is never re-zeroed, and therefore contains information from the complete duration of the simulation).  Periodically, a completely new density of states is generated from the infinite temperature transition probabilities via Eqs.\ \ref{transdoseqn} and \ref{transapprox}\textemdash we call this ``refreshing'' the density of states.  In this way, the WL scheme quickly enforces a broad distribution of macrostates, but the TM refreshing accelerates the convergence of the density of states and hence speeds the modification factor schedule.  In the latter portions of a simulation, the modification factor is so close to one that its effects on $\Omega$ are negligible.  However, the refreshing continuously improves the calculation since the statistics of the $C$ matrix grow better with the number of simulation steps.  At this point, it is even feasible to switch to the bare TM scheme described by the acceptance criterion in Eq.\ \ref{acccriterion2}.  Therefore, the distinction between our approach and the TM method is that we use the WL sampling scheme to guarantee convergence of the infinite-$T$ transition probabilities, whereas the latter method has no such guarantee and might require a good initial estimate  for its transition probabilities \cite{wang3,wang4,wang5}.  

It is important to note that the modification factor schedule in this hybrid scheme might be modified from the original WL approach.  For the Lennard-Jones system, we have found that the $80\%$ ``flat'' histogram requirement for changes to $f$ can be relaxed significantly to the mere requirement that each histogram bin have only one entry.  Furthermore, updates to the modification factor can be more dramatic, such as $\ln{f} \gets 0.1\ln{f}$.  These changes minimize the time that detailed balance is not obeyed, during which entries to the $C$ matrix might be biased due to violation of the microcanonical average in Eq.\ \ref{basiceqn}.  As $\ln{f}$ approaches zero, updates to the infinite temperature transition probabilities grow more accurate.  That is, the faster $\ln{f}$ approaches a near-zero value, the more accurately the density of states estimate will be calculated from the transition probabilities.  One might consider postponing entries in the $C$ matrix and the refreshing of the density of states until the first few stages of the modification factor schedule have passed, during which $f \gg 1$.  It is also possible to refresh only once, at the end of the simulation, which is equivalent to the original WL algorithm with additional data collection (the transition matrix).  These options impart some flexibility to the WL-TM approach, to be tailored to the specific system of interest (for the Lennard-Jones system, which converges relatively fast, we have found little difference among them in our tests).

\section{Test cases}

\begin{figure}
\includegraphics[width=3.375 in]{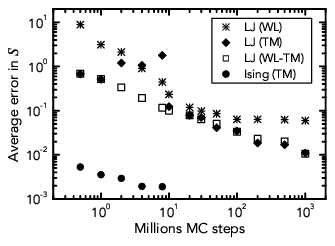}
\mycaption{Error analysis of the dimensionless entropy, $S=\ln\Omega$, as calculated by the Wang-Landau (WL), transition-matrix (TM), and combination (WL-TM) methods for the 110 particle Lennard-Jones (LJ) and 32x32 Ising systems.  Details of each method are described in the text.}
\label{fig:errors}
\end{figure}

We have tested these algorithms on the 110-particle Lennard-Jones system at reduced density $\rho=0.88$.  We cut the potential at $2.5\sigma$ and apply the long-range tail correction \cite{frenkelsmit1}, and we tabulate the transition probabilities for the energy range $-700\epsilon$ to $-500\epsilon$, discretized into 100 bins.   We reject moves which would take the system out of this energy range and accordingly update the transition matrix by $C(I,I) \gets C(I,I)+1$ where $I$ is the initial state; this has the same effect on the transition probabilities and subsequent density of states calculation as if the range were not truncated.  As in the usual implementation of these methods, we calculate the dimensionless entropy, $S=\ln{\Omega}$, rather than the density of states itself.  

We perform six independent runs for each simulation and measure the standard deviation of corresponding entropy bins among them; the statistical error is then measured as the average of this standard deviation over all the bins. (This approach fails to capture systematic errors; therefore, below we also discuss agreement with results from traditional MC simulations.)  For comparison, we also performed the TM method for the 32x32 Ising system, employing single-spin flips and tabulating transition probabilities for all 1023 energy levels.  The estimate of statistical accuracy for the Ising system is the fractional error, $1-\lvert S_\text{run} / S_\text{exact} \rvert$, averaged over each energy level and three independent runs; the exact results are from the method of Ref.\ \cite{beale1}.

\begin{figure}
\includegraphics[width=3.375 in]{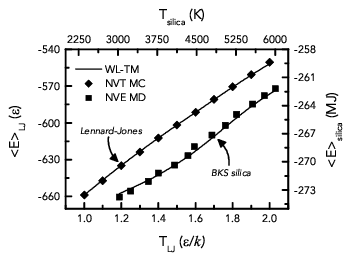}
\mycaption{Comparison of traditional methods (filled symbols) with calculations from density of states generated with the combination (WL-TM) method and reweighted to the appropriate state conditions (lines).}
\label{fig:reweighted}
\end{figure}

The results of these tests are shown in Fig.\ \ref{fig:errors}.  One can see that statistical errors in the WL method rapidly decrease to a limiting value, beyond which additional MC steps do not improve the calculations, as also shown in Ref.\ \onlinecite{jjdp2}.  This is primarily due to the fact that changes in the modification factor eventually become too fast relative to its effect on the density of states.  For the TM method, errors are significant early on in the simulation due to the time needed to generate initial transition probability estimates for each energy level.  Once this range has been covered, the density of states is systematically refined with much greater precision than the WL method.  (Our experience with the TM approach indicates that this initial lag time can grow significantly worse as the energy range broadens, eventually making the TM method infeasible for large entropy gradients.)  The WL-TM algorithm successfully overcomes this drawback and has good convergence properties in both the initial and final stages of the simulation.   

\begin{figure}
\includegraphics[width=3.375 in]{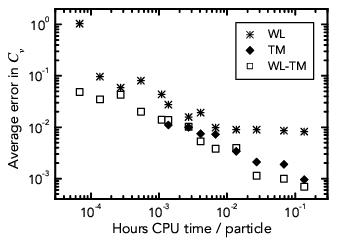}
\mycaption{Errors for each method in the calculated isochoric heat capacity, plotted as a function of CPU time.  For the TM method, results from the early portion of the simulation, during which the relevant energy levels have not yet been visited, are omitted. }
\label{fig:cverrors}
\end{figure}

For the Lennard-Jones system, we have used a standard reweighting procedure \cite{azp1} to generate predictions for the potential energy from the density of states output of the WL-TM algorithm.  In Fig.\ \ref{fig:reweighted}, we show these results alongside data taken from standard $NVT$ MC simulations \cite{nvtcomment}.  The results from the WL-TM algorithm, which are taken only 10 million steps into the simulation, are in excellent agreement with the results of the traditional method.  In Fig.\ \ref{fig:cverrors}, we also show the statistical error in the calculated heat capacity at $T=1.0$, which is measured among results from six independent runs.  The heat capacity results of the WL-TM algorithm appear to have similar convergence behavior as the methods introduced by Yan and de Pablo \cite{jjdp2}, indicating that for simple systems, it does not offer appreciable improvement over these previous approaches.  However, the current method achieves this level of performance without energy derivative information and while remaining readily extendable to any macrostate parameter, including arbitrarily defined ones (e.g., as in extended ensemble formulations).

\begin{figure}
\includegraphics[width=3.375 in]{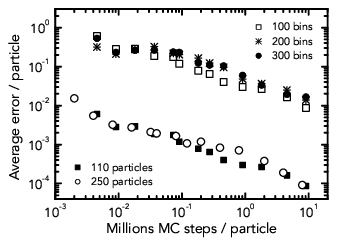}
\mycaption{Effects of system size and energy discretization on the WL-TM method.  All errors are measured for the dimensionless entropy, $S$, and are normalized by the system size.  The upper points, which have been scaled by a factor of 100 for clarity, demonstrate the effects of increasing the energy discretization for a fixed system size of 110 particles.  The lower points correspond to system sizes of 110 and 250 particles at fixed density for a constant degree of discretization ($\sim 2\epsilon$).}
\label{fig:errorsstudy}
\end{figure}

In Fig.\ \ref{fig:errorsstudy} we demonstrate the effects of system size and energy discretization on the WL-TM algorithm.  Results are shown for the original system as well as for a 250 particle system at the same density but with roughly the same energy discretization ($\sim 2\epsilon$).  Here, the statistical error in the dimensionless entropy is normalized by the number of particles, and therefore effectively measures a fractional error.  Fig.\ \ref{fig:errorsstudy} shows that the convergence of the two system sizes is comparable; the 250 particle case does not noticeably suffer from methodological performance losses.  This emerges from the banded nature of the transition matrix: each energy level is effectively connected to only a few energy neighbors due to the small range of energy explored by single particle moves, which have been tuned for 50\% acceptance.  We also investigate the effects of energy discretization on the calculations by increasing the number of energy bins while keeping the system size constant at 110 particles.  Fig.\ \ref{fig:errorsstudy} reveals that fine-graining the energy range only weakly affects the statistical errors, despite the fact that the number of entries in the $C$ matrix (each for which an estimate must be generated) grows as the square of the number of bins.  Presumably this occurs due to the banded nature of the matrix and the subsequent error minimization procedure, which takes into account all transition state pairs in determining the density of states.

We have also investigated our approach on a more challenging system, liquid silica as described by the Van Beest-Kramer-Van Stanten (BKS) potential \cite{silicapotential}.  This system is one for which potential energy derivatives are expensive due to long-range interactions, which encourages methods employing only energy evaluations. The BKS potential treats silica as independent oxygen and silicon ions which interact pairwise through electrostatic and Buckingham-type repulsive/dispersive forces.  Long-range Coulombic interactions are calculated with the Ewald method, using all $k$-vectors with magnitude less than or equal to $5 \times 2 \pi /L$ and $\alpha L = 5.6$.  Short-range interactions are truncated and shifted at 5.5 \AA.   We perform the simulation for 100 oxygen and 50 silicon atoms at a density of 2200 kg/m\textsuperscript{3}, and sample the energy range -5.5 to -5.2 MJ/mol, discretized into 200 bins.  At this density, the dynamics of the silica system are known to experience a dramatic slowdown as the temperature is decreased, which traditionally requires long molecular dynamics trajectories to equilibrate \cite{silicashell}.   We run the proposed WL-TM algorithm for the small silica system for $5 \times 10^8$ MC steps.  For comparison, we conduct multiple NVE molecular dynamics (MD) simulations corresponding to 11 different temperatures, using velocity-Verlet integration with a 1 fs timestep and propagating for $2 \times 10^5$ timesteps.    Fig.\ \ref{fig:reweighted} reveals good agreement between the results of the WL-TM algorithm for silica and those generated from the MD simulations.  This is merely a first step in demonstrating the usefulness of WL-type methods for the glassy silica system, and we leave a more detailed investigation for future work.

\section{Conclusion}

In summary, we have presented a method for calculation of the density of states of a system which combines the good statistical accuracy of transition matrix estimators with the rapid broad sampling of phase space generated by the Wang-Landau method.  This 
implementation requires only potential energy evaluations, and therefore, is general to lattice and continuum systems.  For the same reason and for the additional feature that the transition matrix preserves information from the complete simulation, the method is computationally efficient.  Finally, from our test simulations, our implementation appears to have good convergence properties and allows rapid calculation of the density of states as compared to the original Wang-Landau approach.  The method appears encouraging for studies of complex systems such as dense, supercooled, or glassy liquids, for which gaining accurate results from reasonable simulation times has been challenging.

\begin{acknowledgements}
We gratefully acknowledge the support of the Fannie and John Hertz Foundation and of the Dept.\ of Energy, Division of Chemical Sciences, Geosciences, and Biosciences, Office of Basic Energy Science (grants DE-FG02-87ER13714 to PGD and DE-FG02-01ER15121 to AZP.)
\end{acknowledgements}

\figuresatend


\begin{references}
\bibitem{inftcomment} This derivation and Eq.\ \ref{transdoseqn} also apply to the more general case in which $\paccij$ is any constant.
\bibitem{nvtcomment} These results represent averages over four independent $NVT$ runs for each state point.  In each run, the system is equilibrated for 2.5 million steps before a 5 million step production cycle.
\end{references}
\end{document}